\documentclass[twocolumn,5p,fleqn]{elsarticle}

\usepackage{color}

\usepackage{graphicx}
\usepackage{bm}
\usepackage{amsmath}
\usepackage{natbib}
\usepackage{listings}
\usepackage{hyperref}

\newcommand{\cppVariable}[1]{{\tt #1}}

\newcommand{\cppClass}[1]{{\sffamily #1}}
\newcommand{\cppFile}[1]{\texttt{#1}}

\newcommand{\tildej}{\ensuremath{\tilde{\jmath}}}

\newcommand{\etalter}{{\it et al.\/}}
\newcommand{\eg}{{\it e.\,g.\/}}
\newcommand{\ie}{{\it i.\,e.\/}}

\journal{Computer Physics Communications}

\begin{document}
\setlength{\mathindent}{0cm}

\begin{frontmatter}
\title{Implementation of the SU(2) Hamiltonian Symmetry for the DMRG Algorithm}

\author{G. Alvarez}
\address{Computer Science \& Mathematics 
Division and Center for Nanophase Materials Sciences, Oak Ridge National Laboratory, \mbox{Oak Ridge, TN 37831} USA}

\begin{abstract}
In the Density Matrix Renormalization Group (DMRG) algorithm\cite{re:white92},
Hamiltonian symmetries play an important r\^ole. Using symmetries, the matrix representation of
the Hamiltonian can be blocked. Diagonalizing each matrix block is more efficient
than diagonalizing the original matrix.
This paper explains how the
the DMRG++ code\cite{re:alvarez09} has been extended to handle the non-local SU(2) symmetry in
a model independent way. Improvements in CPU times compared to runs with only local symmetries are
 discussed for the one-orbital Hubbard model, and for  a two-orbital Hubbard model
 for iron-based superconductors. 
The computational bottleneck of the algorithm and the use of shared memory parallelization
are also addressed. 
\end{abstract}

\end{frontmatter}

\section*{PROGRAM SUMMARY}
{\bf Manuscript title:} Implementation of the SU(2) Hamiltonian Symmetry for the DMRG Algorithm\\
{\bf Author:} Gonzalo Alvarez\\
{\bf Program title:} DMRG++\\
{\bf Licensing provisions:} See file LICENSE.\\
{\bf Source code:} \url{http://www.ornl.gov/~gz1/dmrgPlusPlus/}\\  
{\bf Programming language:} C++\\
{\bf Computer(s) for which the program has been designed:} PC\\
{\bf Operating system(s) for which the program has been designed:} multiplatform, tested on Linux\\
{\bf RAM required to execute with typical data:} 1GB (256MB is enough to run included test)\\
{\bf Has the code been vectorized or parallelized?:} Yes\\
{\bf Number of processors used:} 1 to 8 with MPI, 2 to 4 cores with \emph{pthreads}\\
{\bf Keywords:} density-matrix renormalization group, dmrg, strongly correlated electrons, generic programming\\
{\bf PACS:} 71.10.Fd  71.27.+a  78.67.Hc\\
{\bf CPC Library Classification:} 23 Statistical Physics and Thermodynamics\\
{\bf External routines/libraries used:} BLAS and LAPACK\\
{\bf CPC Program Library subprograms used:} None.\\
{\bf Nature of problem:} 
Strongly correlated electrons systems, display a broad range of important phenomena,
and their study is a major area of research in condensed matter physics.
In this context, model Hamiltonians are used to simulate the relevant interactions of a given compound, and the relevant degrees of freedom.
These studies rely on the use of tight-binding lattice models that consider electron localization, where states on
one site can be labeled by spin and orbital degrees of freedom. 
The calculation of properties from these Hamiltonians is a computational intensive problem, since the Hilbert space
over which these Hamiltonians act grows exponentially with the number of sites on the lattice.\\
{\bf Solution method: }
The DMRG is a numerical variational technique to study quantum many body Hamiltonians.
For one-dimensional and quasi one-dimensional systems, the DMRG is able to truncate, with bounded errors and
 in a general and efficient way, the underlying Hilbert space to a constant size, making the problem tractable. \\
%
%
{\bf Running time:} Varies\\

\section{Introduction}

In the DMRG algorithm\cite{re:white92} and other diagonalization-based methods, Hamiltonian symmetries play an important r\^ole.
An operator $\hat{S}$ is a Hamiltonian symmetry if it commutes with the Hamiltonian, \ie, if $[\hat{H},\hat{S}]=0$.
If $S|\psi_1\rangle=s_1|\psi_1\rangle$,
and  $S|\psi_2\rangle=s_2|\psi_2\rangle$, then $\langle\psi_1|H|\psi_2\rangle=0$ provided that $s_1\neq s_2$.
In words, $\hat{H}$ cannot ``connect'' states with different symmetries.
The matrix representation of $\hat{H}$ is then block diagonal, and diagonalizing each matrix block is more efficient 
than diagonalizing the original matrix.

Reference~\cite{re:alvarez09} introduced DMRG++, a generic implementation of the DMRG 
algorithm. There it was shown
 how to take advantage of local symmetries in a generic way, \ie, symmetries $\hat{S}$, 
 such that $\hat{S}=\sum_i \hat{S}_i$, where $\hat{S}_i$ acts only on site $i$.
In this paper the DMRG++ code is extended to handle the non-local SU(2) symmetry.

Many Hamiltonians for strongly correlated electronic systems possess this symmetry, since they conserve 
the full spin. For example, the Heisenberg model with any spin, 
the Hubbard model\cite{re:hubbard63,re:hubbard64b} for any filling with one or multiple orbitals, and the t-J model\cite{re:spalek77,re:spalek07}. 
This is true as long as there are no external magnetic fields.  The implementation of the SU(2) symmetry is involved, particularly if done in 
a generic way, but once implemented it provides substantial performance improvements to each of these models, \eg, the Hubbard model 
with one orbital runs four
times faster for $m\ge 400$, as we will show. All this is achieved without introducing any approximations.

Due to the wide applicability to various models, and the performance improvement that this symmetry brings, it is
studied in detail in this paper, and is implemented in the accompanying DMRG++ code, which
can be found at \url{http://www.ornl.gov/~gz1/dmrgPlusPlus/} .
Section~\ref{sec:dmrgbasis} describes the implementation details of the SU(2) symmetry
for the Hilbert space basis in a model independent way. 
The work on Hilbert space operators is described in section~\ref{sec:operators}, including performance improvements by
using reduced operators with the help of the Wigner-Eckart theorem. The performance bottleneck of the code
is also analyzed, and shared memory parallelization is introduced for the performance critical parts of the code.

In section~\ref{sec:casestudies}, the method is applied first to the Hubbard model and then to a model for iron-based superconductors\cite{re:daghofer08}.
These are new materials whose superconducting pairing mechanism, like in the cuprates, appears to be of electronic origin. 

Finally, a summary is presented. The appendices contain
a few derivations used in the text, as well as some documentation to be able to run the code.

The problem discussed here was treated originally by McCulloch \etalter, in References~\cite{re:mcculloch02,re:mcculloch00}. Comparison
to their results is provided.

\section{Hilbert Space Basis}\label{sec:dmrgbasis}
\subsection{Basis on a Single Site}
Consider the usual\cite{re:cornwell84} SU(2) operators $S^+$, $S^-=(S^+)^\dagger$,  $S^z$, 
and  $S^2=\frac12(S^+ S^- + S^- S^+)+(S^z)^2$. 
In all physical cases these are spin operators--the SU(2) symmetry is actually
a full spin symmetry in the absence of magnetic fields--but this does not concern us at this point.
We consider that the basis is diagonal in $S^2$ and $S^z$, \ie, $S^2|a\rangle = j(j+1)|a\rangle$, and
$S^z|a\rangle=m|a\rangle$. In the code we work with  $\tildej=2j$
instead of $j$, and with $\tilde{m}=m+j$ instead of $m$, because $\tildej$ and $\tilde{m}$ are always non-negative integers.
Since it is standard notation, we will
use $(j,m)$ in the paper; the bijective mapping between $(j,m)$ and $(\tildej,\tilde{m})$ allows us to use $(\tildej,\tilde{m})$
in the code.

We consider that $q$ is the quantum number associated with
some local operator $Q$ (in the case studies it will be the ``total number of electrons'', $N_e$, operator).
We will consider Hamiltonians that conserve $S^2$, $S^z$, and, $Q$. $Q$ can actually be formed by more than one
local operator, using the effective symmetry procedure described in Ref.~\cite{re:alvarez09}.
However, $Q$ should not include $S^z$, that is treated explicitly instead.
In general, $j$, $m$ and $q$, (or equivalently  $\tildej$, $\tilde{m}$, $q$) \emph{does not} 
completely determine the states of the basis. 

In addition to $j$, $m$, and $q$, 
we now introduce a fourth quantum number, \emph{flavor} or $f$, that will be useful in our implementation of
the SU(2) symmetry for DMRG, but $f$ will not necessarily be conserved.
The following definition applies only to states that are eigenstates of $S^2$, \ie, that have a well defined $j$ quantum number.
We define the following relation $|a\rangle\stackrel{f}{\approx}|b\rangle$,
if $\exists\, p\ge 0$  such that either $(S^+)^p |a\rangle = \eta_{p,j,m}|b\rangle$ or 
$(S^+)^p |b\rangle = \eta_{p,j,m}|a\rangle$ holds,
where $\eta_{p,j,m}=\prod_{x=0}^{x=p-1}g_{j,m-x}$ if $p>0$ and  $\eta_{0,j,m}=1$; and
$g_{j,m}=\sqrt{j(j+1)-m(m+1)}$. 
That two states have the same flavor (\ie, that $|a\rangle\stackrel{f}{\approx}|b\rangle$) 
immediately implies that they have the same $j$ value (\ie, that $j_{a}=j_{b}$,
where the notation $j_{a}$
refers to the quantum number $j$ of the state $|a\rangle$, and likewise for $b$).
The relation $\stackrel{f}{\approx}$ is an equivalence relation 
that defines an equivalence class $[|a\rangle]_{\stackrel{f}{\approx}}$ for each element $|a\rangle\in\mathcal{V}$. 
We assign a different non-negative
integer to each equivalence class, and call this number, the \emph{flavor} of that state.

States with the same $f$ and $j$ belong to the same irreducible representation of SU(2). 
That states have the same $j$ does not, \emph{by itself,} imply that they belong to the same 
matrix representing SU(2). For example, in the Hilbert space
of one site with spin 1/2 electrons and a single orbital, the empty state and the doubly occupied state have both $j=0$, but
they do not
belong to the same (one-dimensional) matrix. In other words, these states are not connected by $S^+$. 

Two states have the same triplet $j$, $m$ and $f$, if and only if they are equal (proof in Appendix~\ref{subsec:uniquejmf}).
In other words,  $j$, $m$ and $f$ \emph{completely
and uniquely} determine the
states of the basis.

\subsection{Basis on Multiple Sites: Outer Products}
Consider two vectors spaces with bases $\mathcal{V}_1$ and $\mathcal{V}_2$, respectively. Assume that the
states in these bases are eigenstates of both $S^2$ and $S^z$. Consider the vector space created
by the outer product, $\mathcal{V}_3\equiv\mathcal{V}_1\otimes\mathcal{V}_2$. 
Let $S^+:\mathcal{V}_3\rightarrow\mathcal{V}_3$, be such that $S^+=S^+_1 + S^+_2$ (where the subindices 1 and 2 indicate 
that $S^+_1$ acts only on $\mathcal{V}_1$ and  $S^+_2$ acts only on $\mathcal{V}_2$),
We define $S^z:\mathcal{V}_3\rightarrow\mathcal{V}_3$ and $Q:\mathcal{V}_3\rightarrow\mathcal{V}_3$ in the same way, 
$S^-=(S^+)^\dagger$, and  $S^2=\frac12(S^+ S^- + S^- S^+)+(S^z)^2$.
How can we construct a basis of this outer
product whose states are also eigenstates of $S^2$ and $S^z$?
(One immediately notes that the states $|a\rangle\otimes|b\rangle$, with $|a\rangle\in\mathcal{V}_1$,
and $|b\rangle\in\mathcal{V}_2$, are not necessarily eigenstates of
$S^2$.) The most general solution is
\begin{equation}
|c\rangle=\sum_{a,b}G_{c,a+bN_1}|a\rangle\otimes|b\rangle,
\label{eq:factors}
\end{equation}
where $N_1$ is the number of states in $\mathcal{V}_1$.
In the case of $S^2$ we have an ansatz for $G$ in terms of Clebsh-Gordon coefficients (see, \eg, Ref.~\cite{re:cornwell84}).

Before proceeding to create the basis, we need to explain how to assign quantum numbers to the outer product of states.
It is true that $Q|a\rangle\otimes|b\rangle=(q_a+q_b)|a\rangle\otimes|b\rangle$, and that $S^z|a\rangle\otimes|b\rangle=(m_a+m_b)|a\rangle\otimes|b\rangle$.
But $|a\rangle\otimes|b\rangle$ is not necessarily an eigenvector of $S^2$, as mentioned before.
Therefore, it does not have 
a well defined flavor either. We now extend the definition of flavor for these states in the following way: 
$|a\rangle\otimes|b\rangle\stackrel{f}{\approx}|a'\rangle\otimes|b'\rangle$ if and only if 
all these equalities hold:
$j_{a}=j_{a'}$, $f_a=f_{a'}$, and $q_{a}=q_{a'}$;
$j_{b}=j_{b'}$, 
$f_b=f_{b'}$, and $q_{b}=q_{b'}$. Again, $\stackrel{f}{\approx}$ is an equivalence relation,
and we define equivalence classes, and assign flavors as  different non-negative integers to each equivalence class.
In the few cases where $|a\rangle\otimes|b\rangle$ is an eigenvector of $S^2$, this new definition of $f$ is equivalent
to the previous one.

Then, $|a\rangle\otimes|b\rangle$ is assigned the flavor 
\begin{equation}
\begin{split}
f_{a\otimes b}\equiv &f_a + f_bF_1 + (q_a+q_bQ_1)F_1F_2 + \\
&+(\tildej_a+\tildej_b\tilde{J}_1)F_1F_2Q_1Q_2, 
\end{split}
\label{eq:fatimesb}
\end{equation}
where
$f_a<F_1\,\,\forall |a\rangle\in\mathcal{V}_1$, $q_a<Q_1$, $\tildej_a<\tilde{J}_1$, 
and likewise for $\mathcal{V}_2$. The rationale is similar to the one-site case; states with the same flavor
belong to the same matrix representation of SU(2). In Eq.~(\ref{eq:factors}), the pairs $|a\rangle\otimes|b\rangle$ 
that contribute to a given state $|c\rangle$ all have the same flavor $f_{a\otimes b}$, which 
in turn becomes the flavor of state $|c\rangle$.

Each pair of states\footnote{In the code, these pairs are denoted by a single number, $a+bN_1$.}
 ($|a\rangle$, $|b\rangle$), with $|a\rangle\in\mathcal{V}_1$,
and $|b\rangle\in\mathcal{V}_2$, will contribute to one or more states $|c\rangle$ of $\mathcal{V}_3$.
We first classify the pair $a+bN_1$ in the following way. We calculate all the allowed $j,m$ that
$j_a$, $m_a$ and $j_b$, $m_b$ give rise to. Then, we assign the pair $a+bN_1$ to each one
of these $\mathcal{S}_{j,m,q\equiv q_a+q_b}$ subspaces. After classifying all pairs we
end up with a set of allowed $j,m$ values, and there is one and only one subspace $\mathcal{S}_{j,m,q}$
for each one of those $j,m$ values. The pairwise intersection of these
subspaces is not necessarily empty, because one pair of states $a+bN_1$ may contribute to more than
one state $c$ in Eq.~(\ref{eq:factors}).

Each subspace $\mathcal{S}_{j,m,q}$ is represented
by an object of class \cppClass{JmSubspace}. We now need to determine how many basis states $c$ for $\mathcal{V}_3$ are to be created,
and what  the corresponding factors $G$ are. 
For these tasks, we run the loop given in listing~\ref{lst:createfactors}.
\begin{lstlisting}[caption={Loop that each subspace $\mathcal{S}_{j,m,q}$ of the outer %
product runs to determine the factors $G$ of Eq.~(\ref{eq:factors}).},label=lst:createfactors]
size_t flavorSaved=flavorIndices_[0];
flavors_.push_back(flavorIndices_[0]);
size_t counter=0;
for (size_t k=0;k<indices_.size();k++) {
  if (flavorIndices_[k]!=flavorSaved) {
    flavors_.push_back(flavorIndices_[k]);
    counter++;
    flavorSaved = flavorIndices_[k];
  }
  // G(offset+counter,indices_[perm[k]) =
  // = values_[perm[k]]
  if (heavy_) factors.set(indices_[perm[k]],
    offset + counter, values_[perm[k]]);
}
\end{lstlisting}
In this loop \cppVariable{indices\_} contains the states $a+bN_1$ for this particular $\mathcal{S}_{j,m,q}$, and
\cppVariable{flavorIndices\_} the flavor of each $a+bN_1$ state. For each pair $a+bN1$ we have
computed a vector of \cppVariable{values\_} that contains the Clebsch-Gordan coefficients
$\langle j_a m_a j_b m_b|jm\rangle$. The states of the outer product, $\mathcal{V}_3$,
are labeled here by \cppVariable{offset+counter}, where \cppVariable{offset} is the number of states created by previous subspaces and
\cppVariable{counter} is the number of states created by this subspace. Note that \cppVariable{flavorIndices}  is ordered to simplify the
algorithm, which gives rise to a permutation \cppVariable{perm}. This loop does two main things:
(i) it sets the flavor of each $c$, which is simply the flavor of the $a+bN_1$ values that form part of Eq.~(\ref{eq:factors}),
and (ii) 
it sets \cppVariable{G(offset + counter, indices\_[perm[k]) = values\_[perm[k]]}, as explained before.
Finally, flavors can be reassigned new numbers in the basis $\mathcal{V}_3$.
  
Now we have created a completely (\ie, in $j$, $m$, and $q$) 
ordered basis for $\mathcal{V}_3=\mathcal{V}_1\otimes\mathcal{V}_2$ composed of eigenstates of $S^2$ (and $S^z$ and $Q$).
It is useful to be able to ``disable'' the SU(2) symmetry, 
which is done by just taking $G$ in Eq.~(\ref{eq:factors}) to be the identity, \ie, $G_{c,a+bN_1}=\delta_{c,a+bN_1}$.
We also need a permutation $P^{12}$ to account for effective symmetry ordering\cite{re:alvarez09}.
Then Eq.~(\ref{eq:factors}) becomes
\begin{equation}
|c\rangle=\sum_{a,b}G_{P^{12}(c),a+bN_1}|a\rangle\otimes|b\rangle.
\label{eq:factors2}
\end{equation}
When the SU(2) symmetry is ``enabled'', $G$ is non-trivial and $P^{12}$ is the identity, and vice-versa.
In the DMRG procedure, three types of outer products will appear,
and there will be three factors $G$ and three permutations $P$ at each DMRG step.

The subspaces $\mathcal{S}_{j,m,q}$ for a given outer product space can be heavy or light, and this
is denoted by the \emph{boolean} \cppVariable{heavy\_} in listing~\ref{lst:createfactors}. 
When adding a new site to the system or to the environment, the subspaces are always heavy.
When forming the superblock (by combining system and environment)
the subspaces are heavy if $j=j_{target}$ \emph{and} $q=q_{target}$, and
light otherwise, where $j_{target}$ and $q_{target}$ are the $j$ and $q$ values of the
ground state to be considered by the DMRG algorithm.
Heavy subspaces compute the factors $G$, light subspaces compute only the offsets.
This is done for performance reasons; the factors $G$ are only computed when needed.

\subsection{Change of Basis}
For the DMRG basis transformation the first order of business is to calculate the density matrix for system and environment.
If we label the states with $|j,m,f\rangle$ we get
\begin{equation}
\rho^S_{j_1m_1f_1;j'_1m'_1f'_1} = \sum_{j_2,m_2,f_2}  \psi_{j_1m_1f_1;j_2m_2f_2}^*\psi_{j'_1m'_1f'_1;j_2m2f_2}.
\label{eq:rho}
\end{equation}
One roadblock here is that $\rho^S$ does not necessarily conserve $S^2$ or $S^z$.
To solve this problem, McCulloch {\it et al.} successfully proposed\cite{re:mcculloch02} to use the SU(2) invariant reduced density matrix, 
\begin{equation}
\rho^{S [j_1,m_1]}_{Inv.\,f_1;f'_1} = \sum_{j_2,m_2,f_2}  \psi_{j_1m_1f_1;j_2m_2f_2}^*\psi_{j_1m_1f'_1;j_2m2f_2},
\label{eq:rhoInvariant}
\end{equation}
instead of Eq.~(\ref{eq:rho}), and modify the DMRG truncation procedure accordingly. 

 The DMRG truncation procedure with $\rho_{Inv.}^S$ is as follows. We diagonalize $\rho^S_{Inv.}$, and consider 
 its eigenvectors $W^S$ ordered in increasing eigenvalue order.
Let $m$ be a fixed number that corresponds to the number of states in $\mathcal{V}(S)$ that are to be kept. If $m\ge\#\mathcal{V}(S)$, 
then $W$ remains unchanged. But if $m<\#\mathcal{V}(S)$, then $W$ is truncated by discarding all states above $m$, and thus $W$ becomes a 
rectangular matrix of size $m\times\#\mathcal{V}(S)$. The basis of $\mathcal{V}(S)$ is transformed by applying the (possibly truncated) linear transformation $W^S$.
Operators are transformed in the usual way
$(H^{S {\rm new\,\,basis}})_{\alpha,\alpha'}=(W^S)^{-1}_{\alpha,\gamma} (H^{ S})_{\gamma,\gamma'}W^S_{\gamma',\alpha'}.$
This procedure is repeated for the environment block.

The transformed state $W|\alpha\rangle$ has the same flavor as $|\alpha\rangle$ (see Appendix~\ref{subsec:flavortransform}).
If $|j,m,f\rangle$ is to be discarded, then we need to be sure to
discard all states $|j,m',f\rangle$ for all $m'$, else the remaining basis will not preserve the 
SU(2) symmetry. Alternatively, if $|j,m,f\rangle$ is to be discarded but $|j,m'\neq m,f\rangle$ is not,
then $|j,m,f\rangle$ is kept.

\section{Operators and Optimizations}\label{sec:operators}
\subsection{Product of Operators}
As mentioned before, three types of outer products need be considered: (i) for the outer product of system and a
newly added site(s), (ii) for the outer product of environment and a newly added site(s) and (iii)
for the outer product of system and environment. For the first two the corresponding bases are stored in objects
of class \cppClass{DmrgBasisWithOperators}. For the outer product of system and environment, the outer product is done on-the-fly only because 
of memory storage reasons.
If $A^S$ and $B^S$ are both in the system their product is:  
\begin{eqnarray}
(A^SB^S)_{c,c'}& =& \sum_{a,b,a'}G^S_{P^S(c),a+bN_s} \left(\tilde{s}_a\sum_l A^S_{a,l} B^S_{l,a'}\right)  \times \nonumber\\ 
& & G^S_{P^S(c'),a'+bN_s},
\label{eq:operatorProductSystem}
\end{eqnarray}
where $\tilde{s}_a=(\bar{f})^{n_a}$, $n_a$ is the number of electrons in state $a$, and $\bar{f}= -1$ if $A$ and $B$ anticommute or
$\bar{f}= 1$ if they commute. If $A^S$ is in the system and $B^E$ is in the environment their product is:
\begin{eqnarray}
(A^SB^E)_{c,c'}&=&\sum_{a,b,a',b'}G^{SE}_{P^{SE}(c),a+bN_s} \left(\tilde{s}_aA^S_{a,a'} B^E_{b,b'}\right)  \times \nonumber\\
& & G^{SE}_{P^{SE}(c'),a'+b'N_s}
\label{eq:operatorProductSE}
\end{eqnarray}

\subsection{Wigner-Eckart Theorem and Reduced Operators}
The sums in Eq.~(\ref{eq:operatorProductSystem}) 
and Eq.~(\ref{eq:operatorProductSE}) can be performed in a faster way\cite{re:mcculloch02}
thanks to the Wigner-Eckart theorem.
If operator $A$ transforms as the representation of SU(2) labeled by $J,M$, then 
$\langle f'j'm'| A^J_M |fjm\rangle = C^{j'Jj}_{m'Mm}\langle f'j' || A^J || fj \rangle$, where 
\begin{equation}
\begin{split}
\langle f'j' || A^J || fj \rangle \equiv  & \frac {1}{2j'+1} \times  \\
 &\times \sum_{m',M,m} C^{j'Jj}_{m'Mm} \langle  f'j'm'| A^J_M |fjm\rangle.
\end{split}
\label{eq:reduced}
\end{equation}
Since all operators that appear in constructing the Hamiltonian (for example, $c^\dagger$ in the case of the Hubbard model,
and $S^{z}$, $S^+$ in the case of the Heisenberg model) transform as some representation of SU(2), then
these formulas can always be applied. 

To ``reduce'', for example, Eq.~(\ref{eq:operatorProductSystem}), 
we will first write 
\begin{equation}
G_{c,a+bN_1}=C^{j_c,j_a,j_b}_{m_c,m_a,m_b} \delta_{f_{a\otimes b},f_c},
\end{equation}
where $f_{a\otimes b}$ is given in Eq.~(\ref{eq:fatimesb}), and
then we will gather the sums over $m, M, m'$ together. 

We use throughout the notation $|a\rangle\equiv| f_aj_a m_a\rangle$.
Let us assume that we have calculated the reduced operators
$\langle f_lj_l || A^S || f_aj_a \rangle$ using definition Eq.~(\ref{eq:reduced}),
and similarly for $\langle f_{a'}j_{a'} || B^S ||f_lj_l \rangle$.
We assume that $A^S$ and $B^S$ are both in the system, that they commute (and then $\tilde{f}=1$)
or anticommute (and then $\tilde{f}= -1$), that $\tilde{s}_a=(\tilde{f})^{n_a}$ as before,
that $A^S$ transforms as the irreducible representation of SU(2) labeled by $J_A$ and $M_A$,
and that $B^S$ transforms as the irreducible representation of SU(2) labeled by $J_B$ and $M_B$.

We replace $(A^S)_{a,l}=C^{j_lJ_Aj_a}_{m_lM_Am_a}\langle f_lj_l || A^J || f_aj_a \rangle$, 
and the equivalent for $(B^S)_{l,a'}$ into Eq.~(\ref{eq:operatorProductSystem}).
We obtain an expression for $(A^SB^S)_{c,c'}$ in terms of $\langle j_l f_l|| A^S ||j_a f_a\rangle$
and  $\langle f_{a'}j_{a'} || B^S ||f_lj_l \rangle$.
Then we calculate
$\langle f_{c'} j_{c'} || (A^SB^S) || f_c j_c \rangle$ again using Eq.~(\ref{eq:reduced})
in terms of $(A^SB^S)_{c,c'}$, and replace $(A^SB^S)_{c,c'}$ by the expression we obtained before.
The end result is
\begin{equation}
\begin{split}
&
\langle f_{c'} j_{c'} || (A^SB^S) || f_c j_c \rangle=
\sum_{a_R, b_R, {a'}_R, l_R, J_A, J_B} \mathcal{L}^{S}\,\times\\
&\times \delta_{f_{a\otimes b}f_c}\delta_{f_{a'\otimes b}f_{c'}}
\langle f_lj_l || A^S ||f_aj_a \rangle \langle f_{a'} j_{a'} || B^S ||f_l j_l \rangle \tilde{s}_a,
\end{split}
\label{eq:reducedProductSystem}
\end{equation}
where
\begin{equation}
\begin{split}
 \mathcal{L}^{S}
=&\sum_{m_a,m_a',m_b,m_l,m_c,m_c'}
 C^{j_c,j_a,j_b}_{m_c,m_a,m_b}
 C^{j_{c'},j_{a'},j_b}_{m_{c'},m_{a'},m_b} \times\\ 
& C^{j_l,J_A,j_a}_{m_l,M_A,m_a}C^{j_{a'},J_B,j_l}_{m_{a'},M_B,m_l},
\end{split}
\end{equation}
and $a_R$ represents a sum over $f_a$ and $j_a$ but not over $m_a$, and
likewise for the other indices with subscript $R$.
Note that the factor $\mathcal{L}^S$ depends on $a_R$, $b_R$, ${a'}_R$, $l_R$, $J_A$,
and $J_B$.

We assume now that $B^E$ is an operator in the environment.
Then a similar treatment of Eq.~(\ref{eq:operatorProductSE}) yields:
\begin{equation}
\begin{split}
&
\langle f_{c'} j_{c'} || (A^SB^E) || f_c j_c \rangle=
\sum_{a_R, b_R, {a'}_R, {b'}_R, J_A, J_B} \mathcal{L}^{SE}\,\times\\
&\times \delta_{f_{a\otimes b}f_c}\delta_{f_{a'\otimes b'}f_{c'}}
\langle f_{a'}j_{a'} || A^S ||f_aj_a \rangle \langle f_{b'} j_{b'} || B^E ||f_b j_b \rangle \tilde{s}_a,
\end{split}
\label{eq:reducedProductSE}
\end{equation}
where
\begin{equation}
\begin{split}
 \mathcal{L}^{SE}
=&\sum_{m_a,m_a',m_b,m_{b'},m_c,m_c'}
 C^{j_c,j_a,j_b}_{m_c,m_a,m_b}
 C^{j_{c'},j_{a'},j_{b'}}_{m_{c'},m_{a'},m_{b'}} \times\\ 
& C^{j_{a'},J_A,j_a}_{m_{a'},M_A,m_a}C^{j_{b'},J_B,j_b}_{m_{b'},M_B,m_b}.
\end{split}
\end{equation}

In the code, the class \cppClass{ReducedOperators} keeps track of the 
reduced operators, and the class \cppClass{Su2Reduced} calculates  Eq.~(\ref{eq:reducedProductSystem}) 
and Eq.~(\ref{eq:reducedProductSE}). This results in a substantial speed-up.

\subsection{Shared Memory Parallelization}
The most time consuming part of the DMRG method applied to strongly correlated electronic 
models is the computation of Hamiltonian connections between system and environment.
These connections take the form $c^\dagger_i c_j$ for the Hubbard model,
and $S^+_i S^-_j$, $S^z_i S^z_j$ for the Heisenberg model, and are 
generically represented by Eq.~(\ref{eq:operatorProductSE}) or its reduced form as explained before.
There are a few of these connections in the case of the one-orbital Hubbard model
on a one dimensional chain. There are a few dozen in the case of the two-orbital Hubbard model
for iron-based superconductors on a ladder. 
Then, these connections can be parallelized using, for example, 
\emph{pthreads}\footnote{\emph{Pthreads} or POSIX threads is a standardized C language threads programming interface,
 specified by the IEEE POSIX standard.}, and
the acceleration brought about by this procedure depends on the model, as the results of the next section show.

\section{Case Studies}\label{sec:casestudies}
\subsection{One-orbital Hubbard Hamiltonian}
The one-orbital Hubbard model is given by:
\begin{equation}
H_{U}=\sum_{i,j}t_{i,j}c^\dagger_{i\sigma}c_{j\sigma}+U\sum_{i}n_{i\uparrow}n_{i\downarrow}+\sum_{i\sigma}V_{i\sigma}n_{i\sigma}.
\end{equation}
This model has the SU(2) symmetry 
if we define $S^+=\sum_i c^\dagger_{i\uparrow}c_{i\downarrow}$, $S^z= \sum_i (c^\dagger_{i\uparrow}c_{i\uparrow} - c^\dagger_{i\downarrow}c_{i\downarrow})$,
and $S^2$ as usual from these operators and their transpose conjugates.

We start by reproducing results published in Ref.~\cite{re:mcculloch02} with
$U=1$, $V_{i\sigma} = -0.5$, $t_{ij}=1$ between nearest neighbors, and zero elsewhere, on a 60-site chain at half filling.
These results are shown in Table~\ref{tbl:hubbard60} for $j=0$ and for $j=5$.
The infinite algorithm used $m$ as given in the table, followed by one full sweep with the same $m$.

Having validated these results Table~\ref{tbl:hubbard16} gives additional CPU times for the Hubbard model 
on 16 sites.
In all cases ``Local'' denotes the symmetries $n_e = n_\uparrow + n_\downarrow$ and $s_z =   n_\uparrow - n_\downarrow$,
whereas ``SU(2)'' denotes the symmetries $n_e$, $s_z$ and $s^2$. 
\begin{table}
\centering{
\begin{tabular}{|l|l|l|l|l|}\hline
Symmetry  & m & Energy & CPU   \\\hline
Local & 226  &-76.751582 & 5332  \\\hline
Local & 468  &-76.751733 &86681 \\\hline
Local & 716  &-91.751739 &  320911 \\\hline\hline
SU(2) & 226  &-76.751582 & 1103 \\\hline
SU(2) & 468  &-76.751733& 7564 \\\hline
SU(2) & 716  &-76.751739 & 36640 \\\hline\hline
SU(2) j=5 & 226  &-74.527742 &  1574\\\hline
SU(2) j=5  & 468  & -74.565375& 7188 \\\hline
SU(2) j=5 & 716  & -74.570932&  20364 \\\hline
\end{tabular}}
\caption{\label{tbl:hubbard60} 
Results for the Hubbard model with
$U=1$, $V_{i\sigma} = -0.5$, and $t=1$
on a  60-site chain. Column 2 contains the $m$ total
states kept in each case (this is called $D$ in Ref.~\cite{re:mcculloch02}). 
Energies are in column 3. A factor of $U N /2 = 1\times30/2 = 15$ has been added to all
energies to compare with Ref.~\cite{re:mcculloch02}.
CPU times in seconds are in the last column. 
All rows but the last three refer to the ground-state with $j=0$. 
The last three rows are for the lowest eigenstate with $j=5$.
}
\end{table}
%
%
\begin{table}
\centering{
\begin{tabular}{|l|l|l|l|}\hline
M & SU(2) 1 proc & SU(2) 2 procs & Local 1 proc\\\hline
100 & 42 & 41  & 67\\\hline
200 & 160 & 136 & 319\\\hline
300 & 335 & 290 & 808\\\hline
400 & 544 & 485 & 1602\\\hline
800 & 3020 & 2526 & $>$2 hours\\\hline
\end{tabular}}
\caption{\label{tbl:hubbard16} Times in seconds to run
the one-orbital Hubbard model on 32 sites at half filling, with $U=t=1$.
Runs done with 2 processors used shared memory parallelization with \emph{pthreads}.}
\end{table}

\subsection{Spin 1/2 Heisenberg Model}
This model is given by the Hamiltonian, $\sum_{ij}J_{ij}\vec{S}_i\cdot\vec{S}_j$,
and has full spin symmetry. In this case, and using a 32-site chain with $J_{ij}=1$
only between nearest neighbors, the SU(2) symmetry
yields a speed-up factor roughly between 5 to 10, depending on $m$. However, the shared memory parallelization
performs poorly, because this model has few connections between system and environment blocks.

\subsection{Hamiltonian of Iron-Based Superconductors}
In early 2008, 
 high-temperature superconductivity was discovered\cite{re:kamihara08} in the iron pnictides. 
Except for the cuprates, the iron-based superconductors now have 
the highest superconducting 
critical temperature $T_{\rm c}$ of any material\cite{re:wang08}.
Iron-based superconductors contain conducting layers of iron and arsenic.
As in the cuprate superconductors, in the pnictides there is also evidence that the
superconductivity is not mediated by the electron-phonon interaction\cite{re:yildirim09}, but 
appears to be of electronic origin instead.

A tight-binding two-orbital Hubbard model for the iron pnictides has been proposed\cite{re:daghofer08,re:moreo08}.
This model's kinetic energy is given by
\begin{equation}
K=\sum_{i,\alpha,\gamma,\gamma',\sigma}
t^\alpha_{\gamma,\gamma'} c^\dagger_{i,\gamma,\sigma}c_{i+\alpha,\gamma',\sigma},
\label{eq:feasK}
\end{equation}
where
\begin{equation}
\begin{split}
t^x=\left(
\begin{tabular}{ll}
$-t_1$ & 0\\
0 & $-t_2$
\end{tabular}
\right),\,
%
t^y=\left(
\begin{tabular}{ll}
$-t_2$ & 0\\
0 & $-t_1$
\end{tabular}
\right),\,\\
%
t^{x+y}=\left(
\begin{tabular}{ll}
$-t_3$ & $-t_4$\\
$-t_4$ & $-t_3$
\end{tabular}
\right),\,
%
t^{x-y}=\left(
\begin{tabular}{ll}
$-t_3$ & $+t_4$\\
$+t_4$ & $-t_3$
\end{tabular}
\right).
\end{split}
\end{equation}
The interaction is:
\begin{eqnarray}
H_{int} & = & U_0 \sum_{i\alpha} n_{i,\alpha,\uparrow}
n_{i,\alpha,\downarrow}+\nonumber\\
& + & U_1\sum_{i} n_{i,x} n_{i,y} +  
U_2\sum_{i} \vec{S}_{i,x}\cdot \vec{S}_{i,y}+\nonumber\\
& +  & U_3\sum_{i,\alpha} \bar{n}_{i,\alpha,\uparrow}\bar{n}_{i,\alpha,\downarrow},
\label{eq:feasU}
\end{eqnarray}
where $\bar{n}_{i,\alpha,\sigma}=c^\dagger_{i,\alpha,\sigma}
c_{i,\bar{\alpha},\bar{\sigma}}$ and $\bar{x}=y$, $\bar{\uparrow}=\downarrow$
and $\bar{\bar{a}}=a$.
With this definition, $U_0=U$, $U_1=U'-J/2$, $U_2= -2J$ and
$U_3= -J$. Moreover, usually $U'=U-2J$. 

\begin{figure}
\centering{
\includegraphics{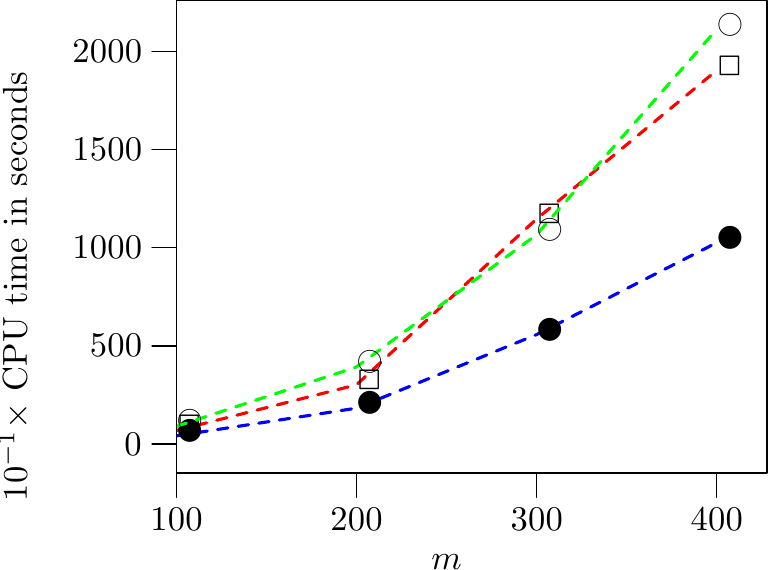}
}
\caption{\label{fig:feas} CPU times in seconds divided by 10,
for the model given by Eqs.~(\ref{eq:feasK}-\ref{eq:feasU}), running on a single core with full spin symmetry (squares),
and with two cores and full spin symmetry (filled circles). The two-core run was done with shared memory parallelization via \emph{pthreads}.
For comparison, the open circles are runs with one core and without the SU(2) symmetry.
All runs were carried out on a 2$\times$4 ladder, with fixed $m=100$ for the infinite algorithm, and with a full finite sweep with the indicated $m$.}
\end{figure}

This model has SU(2) symmetry 
if we define $S^+=\sum_{i,\gamma} c^\dagger_{i\uparrow\gamma}c_{i\downarrow\gamma}$, 
$S^z= \sum_{i,\gamma}( c^\dagger_{i\uparrow\gamma}c_{i\uparrow\gamma} - c^\dagger_{i\downarrow\gamma}c_{i\downarrow\gamma})$,
and $S^2$ as usual from these operators and their transpose conjugates. The sum over $\gamma$ is a sum over the two orbitals, $a$ and $b$ or 0 and 1.
In this model, the efficiency achieved by the use of the SU(2) symmetry is modest. This can be seen, for example, in Fig.~\ref{fig:feas}, by
comparing open circles with squares.
In no case was the gain found to be
larger than a factor of 1.5, and in most cases it was only about 20\% to 30\% depending on $m$ and on the number of lattices sites.

However, the possibility of working with a given total spin ground state facilitates the study of the
nature of ground states. For example, using the SU(2) symmetry it is easier to determine if the ground state is 
a singlet or a triplet. Without the help of the full spin symmetry one would have to run with various $S_z$ target states and
infer from them which one has the lowest energy. 

Using the SU(2) symmetry, the CPU times for this model, which is implemented in class \cppClass{FeBasedSc}, are
given in Fig.~\ref{fig:feas}. The model is expressed on a 2-leg ladder with
parameters\cite{re:xavier09} $t_1=0.058$, $t_2=0.2196$, $t_3=0.20828$, and $t_4=0.079$.
The figure shows the run with a single core and with two cores, parallelized via \emph{pthreads}.
For $m>=300$, CPU times are cut by almost a factor of 2, the theoretical maximum, because
this model, being formulated on a ladder, has many connections, making the shared parallelization efficient.

We end this section on a technical note. 
In this model the real-space basis on a single site has two states that are not eigenstates of $S^2$. 
These states are $|6\rangle \equiv c^\dagger_{\uparrow a}c^\dagger_{\downarrow b}|0\rangle$ and
 $|9\rangle \equiv c^\dagger_{\uparrow b}c^\dagger_{\downarrow a}|0\rangle$. In DMRG++, real-space basis states are 
 coded using a binary number representation, the bit $x$ indicates if there's an electron with internal
 degree of freedom, $x=\gamma+ \sigma N_o$, where $N_o$ is the number of orbitals,
$\gamma$ is the orbital number (0 for $a$ and 1 for $b$), and $\sigma$ is the spin (0 for $\uparrow$ and 1 for $\downarrow$).
For example, $c^\dagger_{\uparrow a}c^\dagger_{\downarrow b}|0\rangle$ has binary number $110$ or 6.

States $|6\rangle$ and $|9\rangle$ are reinterpreted as $1/\sqrt{2}(|6\rangle+|9\rangle)$ and $1/\sqrt{2}(|6\rangle-|9\rangle)$,
respectively. This reinterpretation occurs when calculating operators, such as $c^\dagger_{\sigma\gamma}$,
in this real-space basis, and allows a binary number representation of states to still be used
in this case, even when the original states were not eigenstates of $S^2$.
 
\section{Summary}
By making use of the full spin symmetry to those models that possess it, the DMRG procedure runs faster.
For the one-orbital Hubbard model on a one-dimensional lattice, the speed-up factors were approximately 4 on
a 32-site lattice, and approximately 5 to 10 on a 60-site lattice. All these factors depend on $m$, as detailed in the tables.
The speed-up factor for the two-orbital Hubbard model for iron-based superconductors (\cppClass{FeBasedSc}) on a 2-leg ladder was modest, and
never exceeded 1.5.

The efficiency gained by using the SU(2) symmetry is due to the smaller size of the Hamiltonian matrix
blocks that need to be diagonalized. This effect is countered by the overhead imposed by performing
basis transformations using the factors described in Eq.~(\ref{eq:factors}). However, by employing 
the Wigner-Eckart theorem and using reduced factors and operators, it is possible to bring down the cost of these
transformations significantly. The overall effect is the decrease in CPU times mentioned in the previous paragraph.

Additionally, shared memory parallelization was used to parallelize the calculation of Hamiltonian connections between
system and environment. The success of this method depends on the model, and is most effective when there
are many connections. For the \cppClass{FeBasedSc} model running with 2 cores the speed-up almost reached the theoretical
maximum of a factor of 2. 

Strongly correlated electronic models for iron-based superconductors (implemented in the \cppClass{FeBasedSc} DMRG++ class) is a topic of 
intense study in condensed matter. Of particular interest is the origin and mechanism of the pairing in these superconductors.
The DMRG algorithm provides an accurate way of extracting information from the models 
in this context (for a recent paper, see, \eg, Ref.~\cite{re:berg09}).

DMRG++ is a free and open source implementation of the DMRG algorithm. 
It emphasizes generic programming using C++ templates, friendly user-interface, and as few software dependencies as possible. 
DMRG++ tries to make writing new models and geometries easy and fast by using a generic DMRG engine. 
 
\section{Acknowledgments}
The present code uses part of the psimag toolkit, http://psimag.org/.
I would like to thank Luis G. G. V. Dias da Silva, I. P. McCulloch, M. S. Summers, and J. C. Xavier
for helpful discussions.
This work was supported by the 
Center for Nanophase Materials Sciences, sponsored by the Scientific User Facilities Division, Basic Energy Sciences, U.S.
Department of Energy,  
under contract with UT-Battelle.
This research used resources of the National Center for Computational Sciences, as well as 
the OIC at Oak Ridge National Laboratory.

\appendix

\section{Two states have the same triplet $j$, $m$ and $f$, if and only if they are equal.}\label{subsec:uniquejmf}
Let $|a\rangle$ and $|b\rangle$ be two states with the same $j$, $m$, and $f$.
Without loss of generality we can assume that there $\exists p\ge 0$ such that 
$(S^+)^p |a\rangle = \eta_{p,j,m}|b\rangle$.
Then, because $|a\rangle$ and $|b\rangle$ have the same $S^2$ and $S^z$ eigenvalue, $p$ has to be zero,
implying that $|a\rangle=\eta_{0,j,m}|b\rangle=|b\rangle$.
The reciprocal holds because  a given state has unique values for
$j$, $m$, and $f$. The uniqueness of the first two is trivial.
Flavor is also unique in a given basis, since a state cannot belong to two different equivalence classes.

\section{The reduced DMRG Transformation Conserves Flavor}\label{subsec:flavortransform}
Here we prove that $W|j,m,f\rangle$ has well defined flavor. 
Without loss of generality assume that $(S^+)^p|j,m,f\rangle=\eta_{p,j,m}|j,m+p,f\rangle$.
Since $\rho$ conserves $j,m$, then $W$ does too, and $W|j,m,f\rangle=\sum_{f'}W^{j,m}_{f,f'}|j,m,f'\rangle$,
where $W^{j,m}$ is the matrix block of $W$ corresponding to the good quantum numbers $j,m$. 
Then $(S^+)^pW|j,m,f\rangle=\eta_{p,j,m}\sum_{f'}W^{j,m+p}_{f,f'}|j,m+p,f'\rangle$.
Since the reduced density matrix does not depend on $m$, then nor does $W$. In other words,
$W^{j,m+p}=W^{j,m}$, and so $(S^+)^pW|j,m,f\rangle=\eta_{p,j,m}W|j,m+p,f\rangle$, implying that $W|j,m,f\rangle$ has well
defined flavor. We also proved that $S^+$ and $W$ commute, and since applying $S^+$ does not change flavor and $W$ does
not change $j$ or $m$, then flavors can be assigned in the same way to $W|j,m,f\rangle$ 
as were assigned to $|j,m,f\rangle$. That flavors can be assigned without applying $S^+$
 saves us from keeping track of it through the DMRG procedure.

\section{Building and Running DMRG++}
The required software to build DMRG++ is:
(i) GNU C++, and
(ii) the LAPACK library.
This library is available for most platforms.
The configure.pl script will ask for the \cppVariable{LDFLAGS} variable 
to pass to the compiler/linker. If the \emph{Linux} platform was
chosen the default/suggested \cppVariable{LDFLAGS} will include \cppVariable{-llapack}.
If the \emph{OSX} platform was chosen the default/suggested \cppVariable{LDFLAGS} will
include  \cppVariable{-framework Accelerate}.
For other platforms the appropriate linker flags must be given.
More information on \cppVariable{LAPACK} is here: http://netlib.org/lapack/.

Optionally, make or gmake is needed to use the Makefile, and perl 
is only needed to run the \cppFile{configure.pl} script.

To Build and run DMRG++:
\begin{verbatim}
cd src
perl configure.pl
(please answer questions regarding model, etc)
make
./dmrg input.inp
\end{verbatim}

The perl script \cppFile{configure.pl} will create the
files 
\cppFile{main.cpp}, \cppFile{Makefile} and
\cppFile{input.inp}.
This file can be used as input to run the DMRG++ program. To run the
MPI code the command \verb=mpirun ./dmrg input.inp=
can be used, although  the actual command will vary according to 
the local MPI Installation.

There is also a test suite that can be run for all
standard tests:
\begin{verbatim}
cd TestSuite; ./testsuite.pl --all
\end{verbatim}
or a specific test can be selected and run
by omitting the \verb=--all= argument in the command above. 
Further details can be found in the file README in the code.
\bibliographystyle{elsarticle-num}
\bibliography{thesis}

\end{document}